\documentclass[runningheads, envcountsame, a4paper]{llncs}
\usepackage[utf8]{inputenc}
\usepackage[pdftex]{graphicx}
\usepackage{epstopdf} 
\usepackage{enumitem}
\usepackage{cite}
\usepackage{amsmath,amssymb,amsfonts}
\usepackage{algorithmic}
\usepackage{graphicx}
\usepackage{textcomp}
\usepackage{xcolor}
\usepackage{amsmath}
\usepackage[hyphens]{url}
\usepackage{microtype}
\usepackage{mathtools}

\usepackage{multirow}
\usepackage[ruled,vlined]{algorithm2e}

\DeclarePairedDelimiterX\braket[2]{\langle}{\rangle}{#1 \delimsize\vert #2}

\begin{document}
\title{Quantum Annealing Algorithm for Expected Shortfall based Dynamic Asset Allocation}
\titlerunning{Quantum Annealing for Expected Shortfall}

\author{Samudra Dasgupta and Arnab Banerjee}

\institute{Purdue University}

\authorrunning{S. Dasgupta and A. Banerjee}
\tocauthor{Samudra~Dasgupta and Arnab~Banerjee}
\toctitle{Quantum Annealing Algorithm for Expected Shortfall based Dynamic Asset Allocation}
\maketitle
\setcounter{footnote}{0} 
\begin{abstract}
The 2008 mortgage crisis is an example of an extreme event. Extreme value theory tries to estimate such tail risks. Modern finance practitioners prefer Expected Shortfall based risk metrics (which capture tail risk) over traditional approaches like volatility or even Value-at-Risk. This paper provides a quantum annealing algorithm in QUBO form for a dynamic asset allocation problem using expected shortfall constraint. It was motivated by the need to refine the current quantum algorithms for Markowitz type problems which are academically interesting but not useful for practitioners. The algorithm is dynamic and the risk target emerges naturally from the market volatility. Moreover, it avoids complicated statistics like generalized pareto distribution. It translates the problem into qubit form suitable for implementation by a quantum annealer like D-Wave. Such QUBO algorithms are expected to be solved faster using quantum annealing systems than any classical algorithm using classical computer (but yet to be demonstrated at scale).


\end{abstract}

\section{Introduction}
The 2008 mortgage crisis is an example of an extreme event - also called tail risk. Such crises happen more frequently than what a Gaussian distribution will suggest. Statisticians call this evidence for a fat left tail.

The discipline of extreme value theory tries to estimate such tail risks. Good risk managers realize that risk management should be geared towards extreme events. The day-to-day volatility management is best left to sales departments responsible for profitable growth. In fact well managed institutions excel in maintaining the delicate balance between profitable growth and risk controls (especially on human dimensions). The implication here is that traditional approaches for risk measurement like volatility or even Value-at-Risk are no longer considered good guiding metrics by the Risk division in financial institutions. Instead finance managers increasingly prefer using the expected loss when the  loss has already exceeded a threshold. This is known as Expected Shortfall in literature. To understand why Expected Shortfall is preferred over Value-at-Risk or volatility as risk measure, we direct the reader to the excellent book by McNeil et al.~\cite{mcneil}. Further discussions on this topic can be found in \cite{goldberg37}, \cite{gilli39} and \cite{tasche42}.

Sparse data availablility for tail risks makes estimation of parameters (such as for generalized pareto distribution) in Extreme Value Theory  difficult. The Monte Carlo overhead also becomes significant. Thus a Markowitz type optimization problem that uses Expected Shortfall as the risk measure quickly becomes extremely computationally intensive.

Such a problem is ripe for attack by Quantum Computing. Quantum Annealing and/or universal circuits, through the magic of quantum superposition and entanglement, are anticipated to solve such problems in exponentially less time. QUBO (Quadratic Unconstrained Binary Optimization) problems in particular can be solved faster using quantum annealing systems than any classical algorithm using classical computers.

More generally, much work is going on in leveraging quantum computation for machine learning. See~\cite{riste01} for a discussion on the quantum advantage in data science. \cite{orus02} provides an excellent review of the state of the art in quantum computing for finance while \cite{kopczyk03} provides an easy to follow introduction of quantum computing for data scientists. In risk management domain, \cite{woerner05} provides a quantum amplitude algorithm for measuring Value-at-Risk and Conditional Value-at-Risk. Their calculations prove the theoretical superiority (asymptotically) of quantum algorithms to classical monte carlo approaches.

This paper provides the quantum annealing algorithm in QUBO (Quadratic Unconstrained Binary Optimization) form for a dynamic asset allocation problem using expected shortfall constraint. It was motivated by the excellent work done in~\cite{elsokkary04} which is one of the first applications of D-wave\footnote{See~\cite{peng51},~\cite{long53},~\cite{yin54} and ~\cite{wittek55} for some of the latest implementation examples using the D-Wave machine} based quantum optimizer to a Markowitz setting. However the framing of the problem (e.g. fixed investments associated with fixed assets) made the problem academically interesting but not useful for modern finance practitioners.
 
Section~\ref{sec:PS} states the problem statement, Section~\ref{sec:algo} develops the algorithm step-by-step, Section~\ref{sec:future} lays out potential avenues for future work, and Section~\ref{sec:conclusion} concludes by summarizing the main contributions of this paper. Given the large number of symbols floating around in the paper, we have added an appendix for the reader's ease (comprising definitions of the symbols).

\section{Problem Statement}\label{sec:PS}
Consider a fund manager who has been asked to invest \$1 Mn in the public financial market. She can select from any publicly traded instrument and can change the asset allocation periodically. However, the following four guidelines need to be followed:
\begin{enumerate}
\item Performance will be tied to the absolute return of the portfolio
\item Total risk assumed should be below a threshold
\item Risk threshold should adjust dynamically\footnote{In other words, take lower risk in volatile markets and higher risk in stable times}
\item No shorting\footnote{This means one cannot sell an asset one does not own by temporarily borrowing it from someone and then buying it back again from the market later to return the stock borrowed. The risk of not being able to close a short is often too high during crisis.}
\end{enumerate}

The options for the risk measure can be:
\begin{enumerate}[label=(\alph*)]
\item Volatility ($\sigma$): the standard deviation of portfolio return 
\item Value-at-Risk ($VaR_\alpha$): the VaR at confidence level $\alpha$ is the smallest number $l$ such that the probability that the loss exceeds $l$ is no larger than $(1-\alpha)$
\item Expected Shortfall ($ES_\alpha$): the expected loss given the loss has exceeded $VaR_\alpha$
\end{enumerate}

Both $\sigma$ and $VaR_\alpha$ are unable to capture the tail characterstics needed for extreme value based risk measurement. Modern finance practitioners prefer $ES_\alpha$ which is what we use in this paper. The two questions we need to answer are:
\begin{enumerate}
\item What is the optimal asset allocation at any instant?
\item How should the weighting change with time?
\end{enumerate}

\section{Quantum Annealing for Dynamic Asset Allocation}\label{sec:algo}

\subsection{Input Space}\label{ssec:Inputs}
Here we discuss the input space in more details including some specification choices.

\begin{enumerate}[label=(\alph*)]
\item Asset Universe Composition and the returns matrix $\boldsymbol{R_{NT}}$\\
The main input needed is the matrix of historical asset returns . The matrix has N rows where N denotes the number of assets and T columns where T is the length of the time-series. It is preferable to use ETFs (instead of individual stocks, bonds etc.) to get more diversification benefits with less transaction costs. The $N$ asset ETFs should be spread across asset classes. A good asset universe will include:
\begin{itemize}
\item G20 Currency ETFs
\item Commodity ETFs such as precious metals ETF, agricultural product ETFs (e.g. wheat, cash crops), Utility metals ETF (e.g. copper, aluminium)
\item Real estate ETFs
\item Domestic Stocks ETF (small cap, mid cap, large cap)
\item International Stocks ETF (preferably global market cap weighted ex-US)
\item Bond ETFs, T-bill ETFs
\end{itemize}
\item The vector of asset means $\overrightarrow{R}$ is the mean of $\mathbf{R_{NT}}$ over time dimension. It has $N$ elements corresponding to $N$ assets. Each element of the vector is called $\mu_i$. 
\item The frequency of returns should be daily or higher.
\item The historical time-period covered by the matrix is critical as it must encompass at least two crisis periods e.g. 2002 and 2008 (hence don't choose ETFs which do not have such history).
\item Probability level $\alpha$ (manager choice) determines the probability level for Expected Shortfall calculation. We use $1\%$.
\item Expected Shortfall target at time T, $ES_T$. It serves as the dynamic risk threshold. It increases in magnitude (i.e. risk assumed increases) when market volatility is low and decreases in magnitude (i.e. risk assumed decreases) when market volatility is high. It is quantitatively discussed in the Parameter Initialization section.
\item Variance-covariance matrix $\mathbf{\Sigma} = [\sigma_{ij}]$: It is calculated as:
\begin{equation}
\sigma_{ij}=\frac{ (\overrightarrow{r_i}-\mu_i)\cdot(\overrightarrow{r_j}-\mu_j)}{T-1}
\end{equation}
where $\overrightarrow{r_i}$ and $\overrightarrow{r_j}$ are the asset return vectors (with $T$ elements) for the ith and jth asset respectively. Similarly, $\mu_i = \frac{\sum\limits_{k=1}^{T}{r_{i,k}}}{T}$ and $\mu_j = \frac{\sum\limits_{k=1}^{T}{r_{j,k}}}{T}$ are the asset return means for the ith and jth asset respectively.
\item Precision parameter, L (manager choice): This determines the discretization granularity of the weights vector $\overrightarrow{w}$. It represents the number of qubits used to represent weight $w_i$.
\item Convergence criterion, $\eta \%$ (manager choice): This determines the convergence criterion for the algorithm.
\begin{equation}
\left| 1-\frac{ES_i}{ES_T} \right| \leq \eta\%
\end{equation}
This is needed because the Expected Shortfall form used is an approximation to enable QUBO formulation of the problem for quantum annealing.
\end{enumerate}

\subsection{Output Space} \label{ssec:Outputs}
The output will be a vector of qubits $\overrightarrow{\delta}$ of length $Np$ where N is the number of asset ETFs and $p=\log_2L$. Here L is the precision parameter for weights vector $\overrightarrow{w}$ discussed in the previous subsection~\ref{ssec:Inputs}. The weight $w_i$ of asset i is given by:
\begin{equation}
w_i=\sum\limits_{j=1}^{p}2^{-j}\delta_{ij} \label{w_form}
\end{equation}

Note that $\delta_{ij}$ can be only 0 or 1. Moreover, due to the finite precision effects, $w_i \in [0,1)$ i.e. it can never equal the upper bound 1.

\subsection{Objective Function with Constraints}\label{ssec:InitialObjective}
The essential problem is to maximize $\overrightarrow{w}\cdot\overrightarrow{R}$, where $\overrightarrow{w}$ is the asset weights vector and $\overrightarrow{R}$ is the asset means vector, subject to the following constraints\footnote{Equivalently minimize ($-\overrightarrow{w}\cdot\overrightarrow{R}$)}:
\begin{enumerate}
\item $100\%$ of the amount available is desired to be invested i.e.
\begin{equation}
\sum w_i = 1 \label{constrainti}
\end{equation}
However, due to finite precision effects described in \ref{ssec:Outputs}, the allowed values of $w_i$ are quantized. Thus the best we can do is to push $\sum w_i$ to as close to 1 as permitted. The leftover weight $1-\sum w_i$, if positive, is assumed to sit in cash and not invested. If negative, that small amount is assumed to be borrowed\footnote{This will incur interest expenses but we have ignored that in this iteration}.

\item Consciously limit risk taking i.e.
\begin{equation}
|ES| \leq |ES_T| \label{constraintiii}
\end{equation}

\item No short selling i.e.
\begin{equation}
w_i \geq 0 \label{constraintii}
\end{equation}
This is automatically implied by the form for \eqref{w_form} and hence does not need to be explicitly modeled. 

\end{enumerate}

However this framing is not yet QUBO (remember QUBO stands for quadratic \textbf{\textit{unconstrained}} binary optimization) and hence not suitable for Ising Hamiltonian formulation required by quantum annealers.

\subsection{Quantum Annealing Algorithm}\label{ssec:Objective}
The core of the algorithm is to solve the following optimization problem:
\begin{equation}
    \begin{aligned}
        \min_{w} \quad & \frac{1}{2}w^T C_t w\\
        \textrm{s.t.} \quad & \mu_t^Tw = p \\
                            &\sum{w} = 1, \ w\geq0. 
    \end{aligned}
\end{equation}\\

The algorithm proceeds as follows:

\begin{algorithm}[H]
\SetAlgoLined
\KwResult{W}
 Let $W$ = \{\}\;
 \For{$t\gets 1$ \KwTo $T$} {
    Let $p=\text{mean}(\mu_t)$\;
    Let $EST_t = \frac{\sigma_{spy2008}}{\sigma_{spy\_t}} ES_{spy2008}$\;
    Let $C_t = cov(R_t)$\;
    \While{True} {
        Solve the previous optimization problem for $w$\;
        Let $ESt$ be the mean of the lowest $\alpha$\% values of $w^TR_t$\;
        \uIf{ $\frac{|ESt|}{|EST_t|} > 1 + \epsilon$ } {
            Let $p = p * ( 1 - \delta)$\;
        } \uElseIf{$\frac{|ESt|}{|EST_t|} < 1 - \epsilon$} {
            Let $p = p * ( 1 + \delta)$\;
        } \uElse{
            Add $w$ to $W$\;
            break\;
        }
    }
 }
\caption{Expected Shortfall based Dynamic Asset Allocation}
\end{algorithm}
\vskip 0.5cm

Variables:
\begin{itemize}
    \item $\mu_t$: the mean return vector of time window $t$
    \item $\sigma_{spy\_t}$: the volatility of SPY's returns during time window $t$
    \item $C_t$: the co-variance matrix of the sample returns during time window $t$
    \item $\alpha$: risk level parameter
    \item $EST_t$: the target expected shortfall at time window $t$
    \item $ESt$: the expected shortfall during the optimization process at time window $t$
    \item $\epsilon$: error tolerance parameter
    \item $\delta$: profit adjustment scale, could change dynamically during runs
\end{itemize}

\section{Future Work}\label{sec:future}
Quantum Computing has a lot of potential applications in financial risk management. There is a plethora of NP hard optimization problems in a wide varierty of sub-fields such as liquidity risk management, credit risk management, operational risk management etc.

The immediate domains of future work for the problem we have discussed in this paper are:
\begin{itemize}
\item Implement the same problem in a universal circuit model quantum computer.
\item Compare the performance of D-Wave vs IBM-Q for time taken, solution quality, robustness to noise and ability to scale.
\item Use base-N ($N>2$) encoding to reduce the number of qubits required. For this we need to use physical quantum systems that can have more than two eigen states for measurement.
\item On the purely financial optimization front, the model can be made more realistic by adding liquidity constraints on the ETFs, removing no short selling constraints and implementing a regime switching model for volatility (used for setting the Expected Shortfall target). However, these non-critical refinements can await adequate qubit capacity.
\end{itemize}

\section{Conclusion}\label{sec:conclusion}
To our knowledge, this is the first attempt at a quantum algorithm for an Expected Shortfall based Dynamic Asset Allocation. More specifically, a QUBO (or Ising Hamiltonian) formulation for the Expected Shortfall Optimization problem is provided.

This is significant because modern finance practitioners have very little faith on volatility or Value at Risk as an efficient risk metric and so approaches like~\cite{elsokkary04} will find limited adoption. The current management philosophy on risk management is geared towards extreme tail risks (occurrence frequency once in a decade or less) and not towards day-to-day business-as-usual risk.

The algorithm is dynamic because the risk target emerges naturally from the market volatility. There is no arbitrary (expert judgment based) input from a portfolio manager. We have not come across a dynamic quantum annealing algorithm for financial risk management in literature.

Moreover, the algorithm avoids complicated statistics like generalized pareto distribution (GPD)~\cite{mcneil}. GPD requires sufficient tail data for estimating parameters. But if there were sufficient data, then GPD would not have been needed! So it is a tough balancing act and the implementation often boils down to judgement driven heuristics. We avoid this problem by using a simple approach (quadratic in weights) that uses a gaussian as an initial crude guess for the tail and then iteratively converges on the true Expected Shortfall. This approach has the added advantage of avoiding two pitfalls that practitioners often face:
\begin{itemize}
\item deriving risk measures from a time series that has been generated using monte carlo simulation (that used a probability distribution fitted over sparse data) OR
\item undermining risk by using the normality assumption shortcut
\end{itemize}

Lastly, an interesting feature of our algorithm is focus on asset class diversification. This is important because in calm markets some pairs of instruments within the same asset class (e.g. JPM vs Caterpillar within the asset class `stocks') may show misleadingly low correlation. But that diversification benefit vanishes in times of extreme volatility (e.g. everything heads downwards - there is no safe harbor) - exactly when you desperately need the benefit to kick in! We use a small number of ETFs but cover asset classes more comprehensively thereby spanning a broader financial universe which generates superior efficient frontiers.

\section*{Appendix}
Convention followed: Matrices are in capital, bold e.g. $\mathbf{R}$. Vectors are in small, with an arrow over e.g. $\overrightarrow{w}$. Numbers are in small e.g. $\sigma$. Below is a list of the variable definitions:

\begin{itemize}
\item N: number of assets
\item T: number of data points in the time series for each asset
\item $\mathbf{R_{NT}}$: asset returns matrix. The matrix has N rows where N denotes the number of assets and T columns where T is the length of the time-series.
\item $\overrightarrow{R}$: mean (over time) asset return vector with $N$ elements
\item $\mu_i$: ith member of $\overrightarrow{R}$
\item $\overrightarrow{r_i}$: asset return vector (with $T$ elements) for the ith asset
\item $\overrightarrow{w}$: weights vector
\item $L$: number of qubits used to represent weight $w_i$
\item $p$: = $\log_2L$
\item $\mathbf{\Sigma}$: variance-covariance matrix  
\item $\sigma_{ij}$: an element of the variance-covariance matrix
\item $\sigma$: volatility
\item $\alpha$: confidence Level
\item $VaR_\alpha$: value-at-risk at confidence Level $\alpha$
\item $ES_\alpha$: Expected Shortfall at confidence Level $\alpha$
\item $ES_i$: portfolio's expected shortfall at iteration i
\item $ES_T$: portfolio's expected shortfall target (for current time T)
\item $\eta$: convergence threshold
\end{itemize}

\section*{Acknowledgment}
We thank Peter Lockwood for the valuable suggestions and edits.

\end{document}